\documentclass[ aip,
 amsmath,amssymb,
preprint,%
]{revtex4-1}

\usepackage{graphicx}
\usepackage{dcolumn}
\usepackage{bm}
\usepackage{array}
\usepackage{longtable}
\usepackage{braket}
\usepackage{comment}
\usepackage{hyperref} 

\usepackage[utf8]{inputenc}
\usepackage[T1]{fontenc}
\usepackage{mathptmx}
\usepackage{etoolbox}
\makeatletter
\def\@email#1#2{%
 \endgroup
 \patchcmd{\titleblock@produce}
  {\frontmatter@RRAPformat}
  {\frontmatter@RRAPformat{\produce@RRAP{*#1\href{mailto:#2}{#2}}}\frontmatter@RRAPformat}
  {}{}
}
\makeatother
\begin{document}


\title{Extending the definition of atomic basis sets to atoms with fractional nuclear charge}
\author{Giorgio Domenichini}
 \email[]{giorgio.domenichini@univie.ac.at}
 \affiliation{University of Vienna}

\date{\today}

\begin{abstract}
Alchemical transformations showed that perturbation theory can be applied also to changes in the atomic nuclear charges of a molecule. The alchemical path that connects two different chemical species involves the conceptualization of a non-physical system in which atom possess a non-integer nuclear charge. A correct quantum mechanical treatment of these systems is limited by the fact that finite size atomic basis sets do not define exponents and contraction coefficients for fractional charge atoms. This paper is proposes a solution to this problem, and shows that a smooth interpolation of the atomic orbital coefficients and exponents across the periodic table is a convenient way to produce accurate alchemical predictions, even using small size basis sets.
\end{abstract}
The following article will be submitted to AIP - The Journal of Chemical Physics 

\maketitle

\section{\label{sec:introduction}Introduction}

A molecular compound can be defined by its composition, its molecular geometry, and its molecular charge; in a computational simulation, these properties are fully described by: nuclear positions $\mathbf{R}$, nuclear charges $\mathbf{Z}$, and number of electrons $\mathrm{N}_e$. \\
Considering only compounds in their electronic ground state, and assuming valid the Born-Oppenheimer adiabatic approximation\cite{Born_Oppenheimer},  $\mathbf{R}$, $\mathbf{Z}$,  and $\mathrm{N}_e$ are the variables which define a continuous P.E.S. (Potential Energy Surface):  $E(\mathbf{R},\mathbf{Z},\mathrm{N}_e)$.
The derivatives of the energy with respect to these coordinates give many insights about chemical properties and chemical reactivity.\cite{cardenas2011,cardenas2020nuclearfukuiforces_ayes,cardenas2016chem_pot,gomez2021alchhardness_cardenas,Cohen_Pirovano1994_nuclear_fukui,Cohen_Pirovano1995_nuclear_fukui,Balawender2001nuclearfukui,Moreno_ADFT,Balawender1998Fukui,baekelandt1996nuclearFukui,robles2018localelectroph,geerlings2019new}

Perturbations along the nuclear charge coordinates, meaning changing the chemical composition of one reference molecule to one or more target molecules, are known as alchemical transmutations.\cite{Cardenas2020Deprotonation,munoz2017predictive,BALAWENDER202315,Keith2019,Keith2017,Keith2020,Griego_Keith2021_comp_guidelines,Keith2021QML,guido2021AlchChirality,von2007alchemical,apdft,Lilienfeld2009,Rudorff2019a,Rudorff2020,von2006molecular,marcon2007tuning,krug2022relative,krug2023generalized,von2023even}
Perturbation theory can be applied to alchemical transmutations, but it requires the abstraction of calculating derivatives along a non-physical path. The concept of nuclear charge has to be extended to non-integer numbers: if the basis set is kept fixed, a molecule with fractional nuclear charge can be modelled by scaling the nuclear-electron attraction operator by a real number.

Contracted Gaussian type orbitals (CGTO)\cite{Clementi1965_atomic,CLEMENTI1966223} are standard atomic basis for quantum mechanical calculations on molecular systems, their contraction coefficients and exponents are optimized and tabulated for every chemical element. 

Previous works\cite{balawender2019exploringGeerlings,Domenichini2020} showed that the alchemical perturbation (AP) series calculated using the atomic orbital (AO) basis set of the reference, converges to the energy of the target molecule with the basis set of the reference ($E^{T[R]}$). The difference between the target's energy calculated with reference basis set, and the target's energy with its own basis set($E^{T[T]}$), was defined\cite{Domenichini2020} as "alchemical basis set error" $\Delta E_{BS}=E^{T[R]}-E^{T[T]}$.  
Using small basis sets, $\Delta E_{BS}$ can be as big as 1. Ha for the transmutation of a just one atom. Such huge error makes meaningless any alchemical energy prediction for basis sets such as STO-3G,STO-6G\cite{stos}, Pople spilt valence basis sets 3-21G, 6-31G\cite{321g,631g}, Dunning's cc-pVDZ and cc-pVTZ\cite{Dunning_1989,Dunning_1993}, and Ahlrichs's def2-TZVP\cite{def2tz}.

A way to reduce this error is to use of large augmented basis sets \cite{keith_22_QA_cycles,keith_22_QA_diatomics_H,Chang_bonds} (e.g. the Dunning's aug-cc-pVQZ, aug-cc-pV5Z ).
Also uncontracted basis set are more flexible than their respective contracted versions (e.g. for uncontracted cc-pVDZ \cite{anm} uncontracted Cartesian Def2TZVP \cite{Chang_bonds}), therefore more suitable to applications in quantum alchemy.
According to a previous investigation\cite{domenichini2022alchemical} we found that the smallest alchemical basis set errors are produced using polarized core and valence basis functions, such as Dunning's cc-pCVnZ\cite{woon1995a}, or Jensen's pC-n \cite{pCn_jensen2001a,pCn_jensen2002a,pCn_jensen2007a} basis sets, similar conclusions were obtained also by Geerlings \textit{et al.}\cite{geerlings2019new}. Particular good results can be achieved using the Jensen's pcX-n \cite{pcXbs} basis sets. 
Some basis set do not differ from reference to target, this is the case of the universal Gaussian basis set\cite{ugbs}, as well as the plane waves basis sets.\cite{plane_waves,blochl1994}  

The most elegant way to solve the basis set issue, suggested in a 2012 paper from M.Lesiuk, R.Balawender and J.Zachara \cite{zachara2012Geerlings}, involve a smooth transformation of the atomic orbitals as the nuclear charges of the reference approaches the nuclear charges of the target molecules. 
In this paper I show that, as long as target and reference molecules share the same number of primitive and contracted Gaussian basis function, it is straightforward to interpolate through Splines basis set exponents and coefficients, extending the definition of basis sets to atoms with a non integer nuclear charge. The extended basis set that change consistently with the nuclear charge will be called alchemically consistent basis sets (ACBS), a term which is not referred to a particular basis set, rather to the interpolating procedure on which can be derived from an existing basis set.
\begin{figure}[h!]
    \centering
    \includegraphics[width=.5\linewidth]{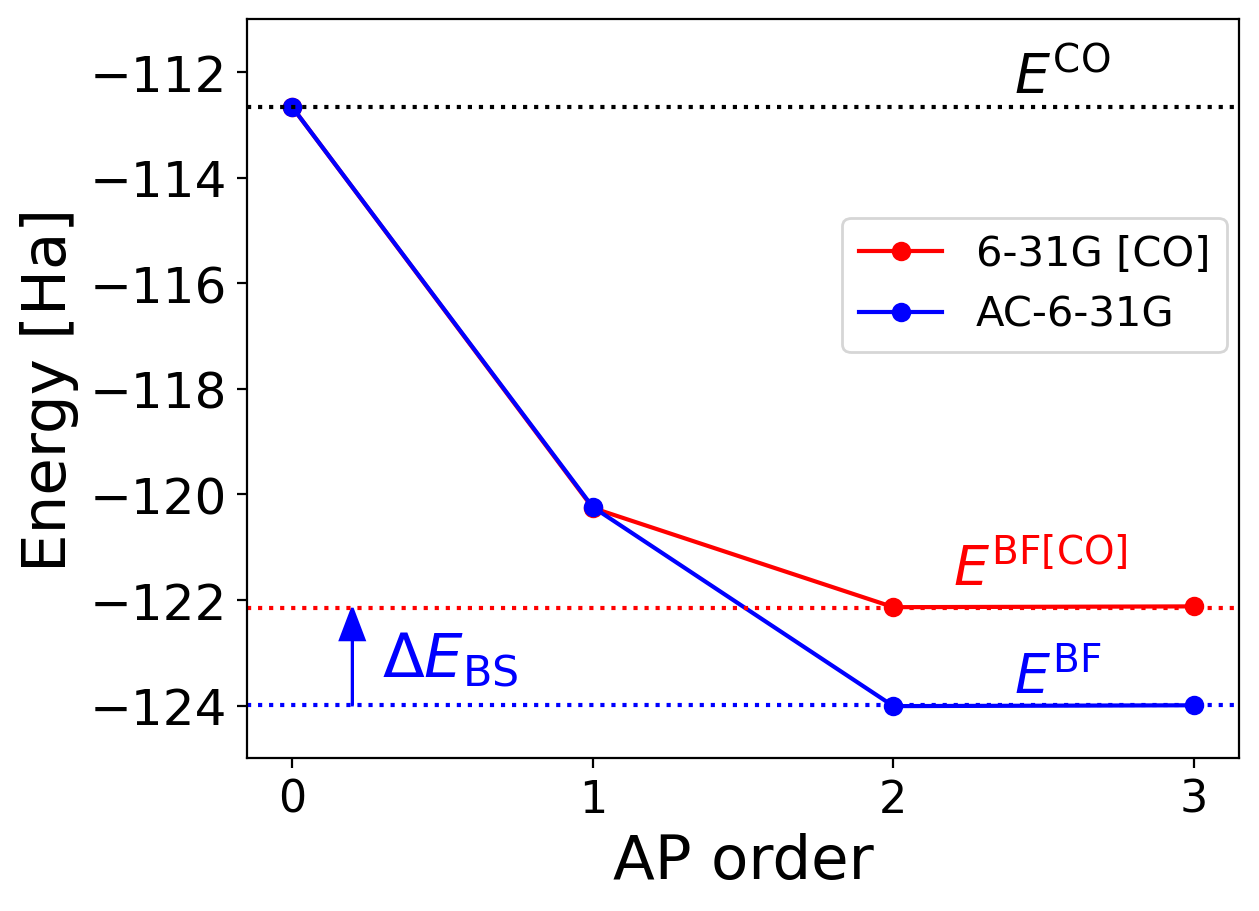}
    \caption{For the simple alchemical transmutation CO$\rightarrow$BF (HF/6-31G), using reference basis set, the AP series converges $E^{BF[CO]}$, while using AC-6-31G basis set the series approaches $E^{BF[BF]}$}
    \label{fig:abse}
\end{figure}
In section \ref{sec:Methods} are explained the interpolation procedure through alchemical consistent basis sets are defined. For the Restricted Hartree Fock (RHF) level of theory is proposed a formula to calculate the first alchemical derivative including the basis set derivatives.
In section \ref{sec:num_ex} are given some example on how including basis set derivatives in the alchemical expansion is crucial to produce accurate results while using a small basis set. Numerical evidence of this are deduced from the two test cases of diatomic molecules (exemplified in Figure \ref{fig:abse}) and the boron-nitrogen (BN) doping of benzene.

The results in this paper were obtained from a locally modified version of the PySCF program\cite{pyscf_article,pyscf_2020recent}, accessible through the GitHub repository "Supplementary code for Quantum Alchemy"\cite{Supplementary_QA}.
In this work were used Python\cite{van1995python,ipython} 3.8 scripts, using Numpy\cite {harris2020numpy}, Scipy\cite{2020SciPy-NMeth} and Matplotlib\cite{hunter2007matplotlib} libraries.

\section{\label{sec:Methods} Methods}
\subsubsection{\label{sec:spline}Interpolation of orbitals exponent and coefficients}

The radial part of a CGTO is:
\begin{equation}
    R^{CGTO}(r)= N \sum_i c_i \left({\frac {2\zeta _{i}}{\pi }}\right)^{{3/4}}e^{-\zeta_i r^2}
\end{equation}

N is a normalization constant chosen such that the atomic orbitals are normalized, this constant is calculated automatically by PySCF.

The contraction coefficients $c_i$ and the Gaussian exponents $\zeta_i$, are tabulated in the basis set for different elements. 
In many basis sets, elements belonging to the same period share the same number of atomic orbitals and Gaussian primitives.
This allows a smooth transformation of orbital coefficients and exponents from one element to another. Coefficients $c_i(Z)$, and exponents $\zeta_i(Z)$ have to be defined as continuous and differentiable functions of the atomic nuclear charge $Z$, and for integer nuclear charge they should be equal to the tabulated value.

\begin{figure}[ht]
    \centering
    \includegraphics[width=.4\linewidth]{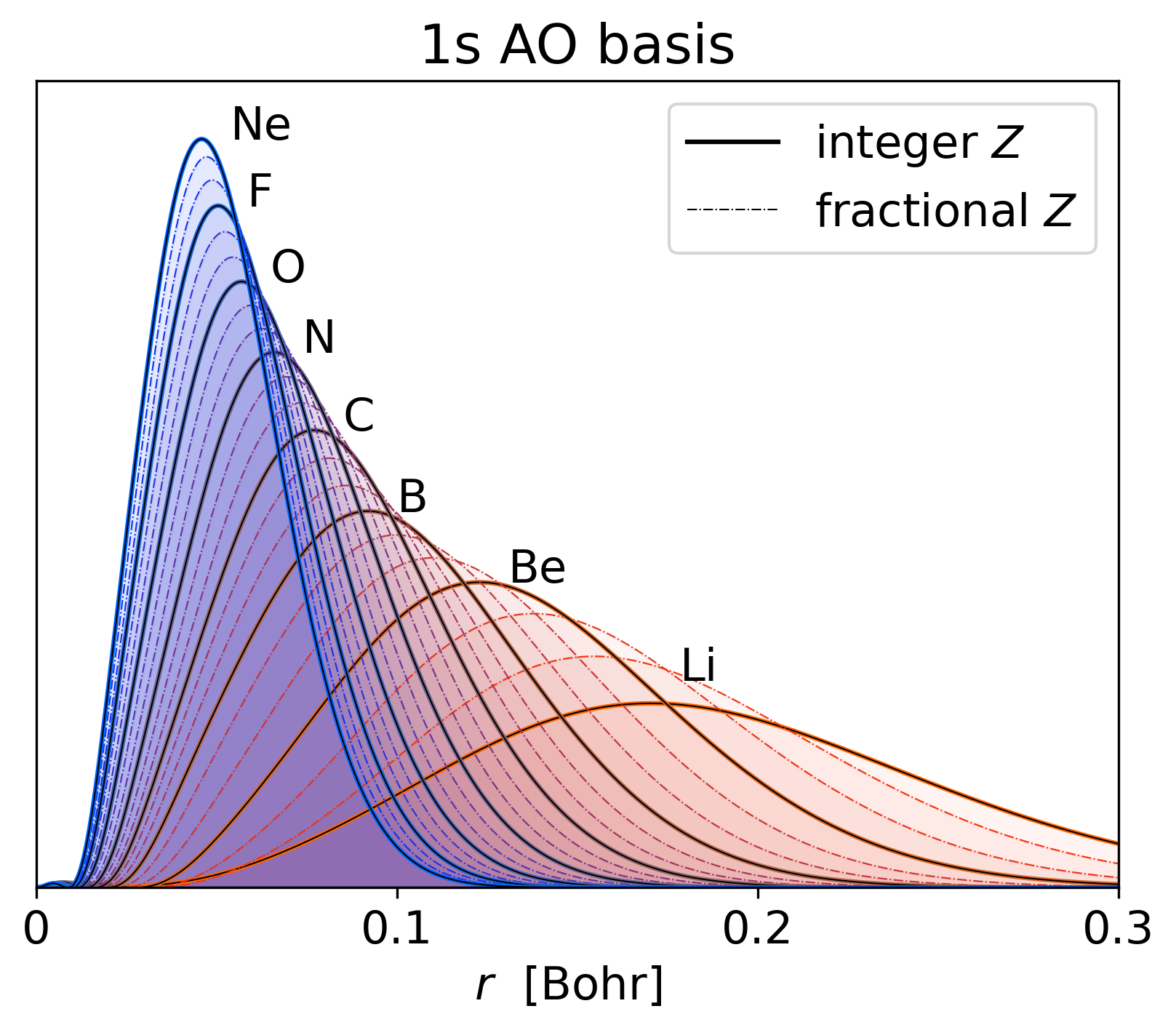}
    \includegraphics[width=.4\linewidth]{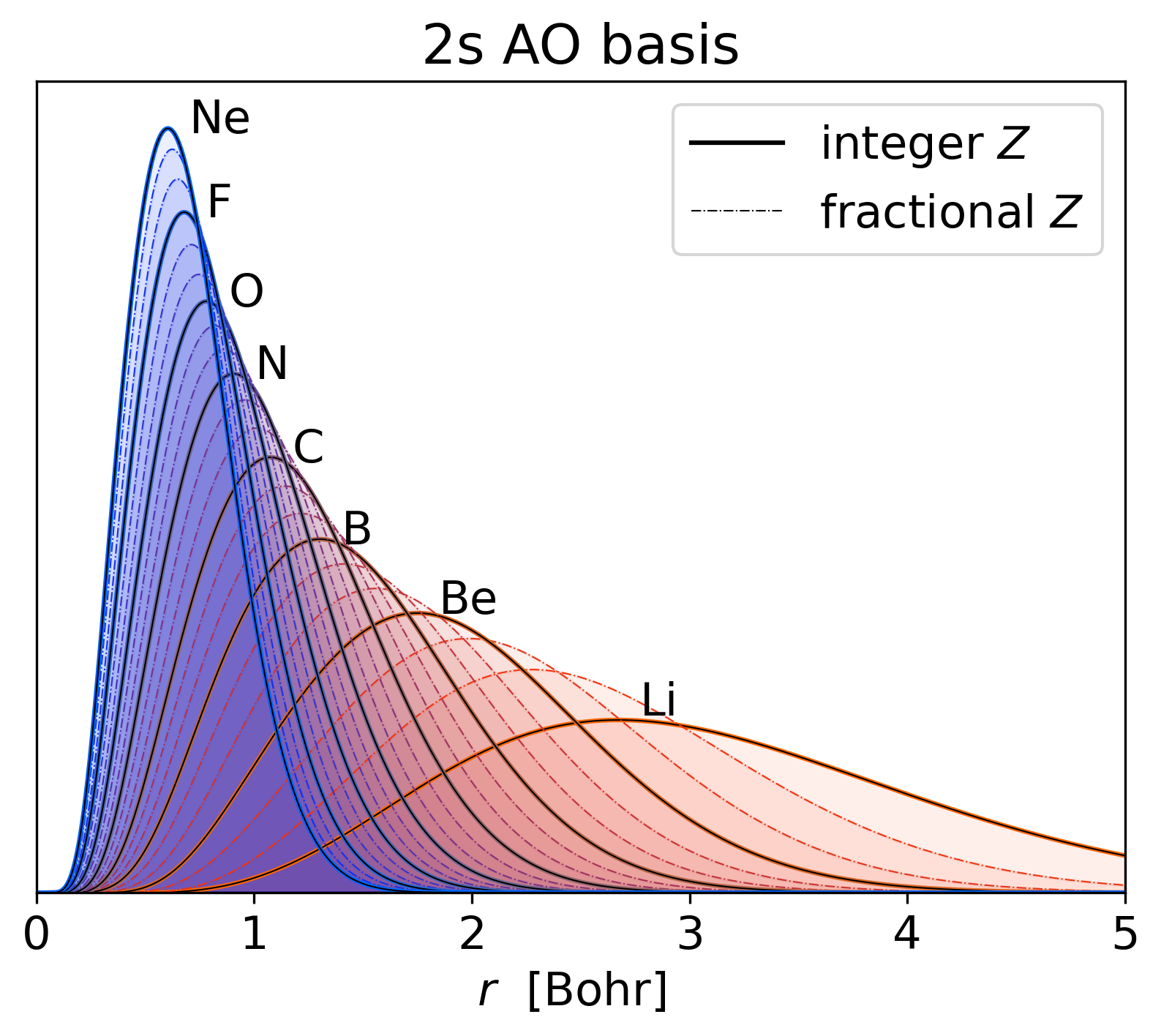}
    \caption{\label{fig:1s2s_orbitals} Radial density function of Pople's 6-31G $1s$ and $2s$ atomic orbitals for the elements of the second period. The solid lines represent the basis function of the elements, which have an integer nuclear charge $Z$, the dashed lines represent the basis functions of some non-physical atoms with fractional nuclear charge $Z+ \frac{1}{3}$ and $Z+ \frac{2}{3}$ }  
\end{figure}

A simple choice that fulfills these requirements is to define $ c_i(Z),\zeta_i(Z) $, is to use splines interpolations. In this paper will be used a fifth order spline with the ‘not-a-knot’ boundary conditions. In Figure \ref{fig:1s2s_orbitals} are plotted, for Pople's 6-31G basis set, the tabulated a, with integer $Z$, and interpolated a with fractional $Z$.

\begin{figure}[h!]
    \centering
    \includegraphics[width=.4\linewidth]{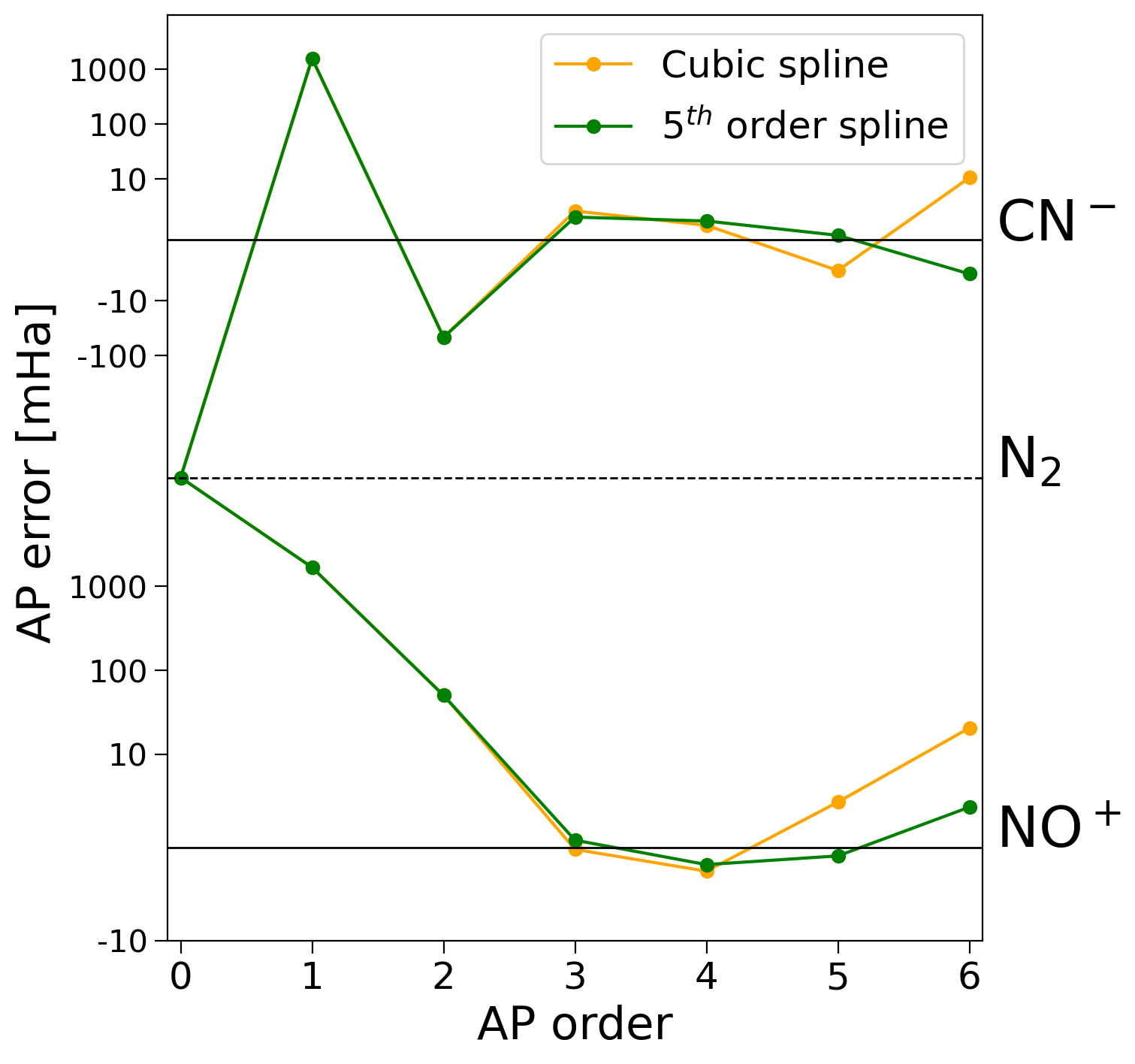}
    \caption{\label{fig:Spline_order} RHF/6-31G alchemical energy predictions: N$_2\rightarrow$CN$^-$ , N$_2\rightarrow$NO$^+$ . Derivatives are calculated numerically from a 7 point central finite difference stencil, with spacing $h=0.05 e$. ACBS are defined using either a cubic spline or a 5$^{th}$ order spline.}
\end{figure}

A $n^{th}$ order spline is a polynomial function that is continuous up to its $n-1^{th}$ derivative. Discontinuity of higher order derivatives affects the alchemical perturbations. As an example, Figure \ref{fig:Spline_order} shows that if AO coefficients and exponents are transformed with a cubic spline, the alchemical series diverges for a perturbation order greater than 4, while if the AO coefficients and exponents transformed with the fifth order spline diverges for an alchemical perturbation order greater than 6.

As a rule of thumb, the AO interpolation spline should have an order which is higher than the alchemical perturbation.

\subsubsection{\label{sec:analyticalDeriv}First order analytical alchemical derivative}
The first order alchemical derivative of the RHF energy can be derived from the gradient expression\cite{pople1979derivative,MOCCIA1970260,Bishop66_RHFgrad}. 
\begin{equation} 
\label{eq:AD1}
\frac{\partial E}{\partial Z}= \sum_{\mu\nu}P_{\mu\nu}\frac{\partial h^{(1)}_{\mu\nu}}{\partial Z}+\frac{1}{2}\sum_{\mu\nu\lambda\sigma}
P_{\mu\nu}P_{\lambda\sigma}\frac{\partial}{\partial Z}(\mu \lambda | | \nu\sigma)+\frac{\partial V_{nuc}}{\partial Z} 
-\sum_{\mu\nu}W_{\mu\nu}\frac{\partial S_{\mu\nu}}{\partial Z}
\end{equation}
$W$ is the energy weighted density matrix:
\begin{equation} W_{\mu\nu}= \sum_i ^{mo.occ.} \epsilon_i c_{\mu i} c_{\nu i}^\dagger
\end{equation}

If the alchemical derivatives are calculated using the reference basis set, the derivatives of the AO integrals w.r.t. the nuclear charge are zero $\left( \frac{\partial }{\partial Z} =0 , \frac{\partial}{\partial Z}(\mu \lambda | | \nu\sigma) = 0 \right)$, and Equation \ref{eq:AD1} is reduced to the Hellmann-Feynman derivative:
\begin{equation}
\label{eq:HFD1}
\frac{\partial E}{\partial Z}= \sum_{\mu\nu}P_{\mu\nu}\frac{\partial h^{(1)}_{\mu\nu}}{\partial Z}+\frac{\partial V_{nuc}}{\partial Z}= \int \rho(\vec{r}) \Delta v (\vec{r}) d \vec{r}+\frac{\partial V_{nuc}}{\partial Z}
\end{equation}

Higher order alchemical derivatives can be evaluated analytically from the solution of the Coupled Perturbed Hartree Fock (CPHF), or the Couple Perturbed Kohn Sham (CPKS) equations \cite{cave1969CPHF,dalgarno1962cphf,cammi_pcm}. These analytical derivatives have been implemented and used in alchemical perturbations \cite{balawender2019exploringGeerlings,c60_2018Geerlings,benzene2013Geerlings, Balawender2018,domenichini2022alchemical}, but only in the case where basis sets do not depend on nuclear charge (reference basis set are kept fixed), and will be used in the following section.
An implementation of analytical CPHF second and third derivatives which includes the basis set derivative (ACBS) is possible, but not yet implemented.
Second and third alchemical derivatives, using ACBS, were calculated from numerical differentiation (through a three point central finite difference scheme with spacing $h=0.05 e$) of the first alchemical derivative (Equation \ref{eq:AD1}).  

\section{\label{sec:num_ex} Numerical examples}
The basis sets 3-21G, 6-31G, cc-pVDZ have a constant number of CGTO and Gaussian primitives for all elements of the second period, this allows to define ACBS which connect all atoms from nuclear charge 3 (Lithium), up to nuclear charge 10 (Neon).
The RHF energy alchemical prediction calculated using the ACBS will be compared to the AP calculated on the basis set of the reference molecule.

\begin{table}[h]
    \centering
    \begin{tabular}{c|c|c}
Basis set & MAE AP3 ACBS & MAE AP3 Ref.BS \\
\hline
 3-21G    &  0.0077  &   1.2653 \\
 6-31G    &  0.0039  &   1.2701 \\
 cc-pvDZ  &  0.0055  &   1.2642 \\
    \end{tabular}
   \caption{MAE of the third order alchemical perturbation of the 14 electrons diatomic molecules (BF,CO,N$_2$), all target within a $\Delta Z = \pm 1$ perturbation are reached, and the interatomic distance is kept fixed at a length of 2.05 Bohr}
    \label{tab:diatomics}
\end{table}

In table \ref{tab:diatomics} are listed the mean absolute errors (MAE) of the third order alchemical prediction (AP3) of the 14 electrons diatomic molecules BF,CO,N$_2$. From the reference molecule are reached all possible target with a $\Delta Z = \pm 1$ perturbation:
\begin{itemize}
    \item N$_2$ $\rightarrow$ NO$^+$ ,CN$^-$ ,CO 
    \item CO $\rightarrow$ CF$^+$, NO$^+$,  BO$^-$, CN$^-$ , N$_2$, BF 
    \item BF $\rightarrow$ BNe$^+$, CF$^+$, BeF$^-$, BO$^-$,CO,BeNe 
\end{itemize}

The error made using the ACBS, is on the milli Hartree scale, and it comes from the truncation of the alchemical series, while the MAE from the reference basis set AP it is constantly off by the huge amount of $\approx 1.2$ Hartree, mainly due to $\Delta E_\text{BS}$. 
\begin{figure}[h!]
    \centering
    \includegraphics{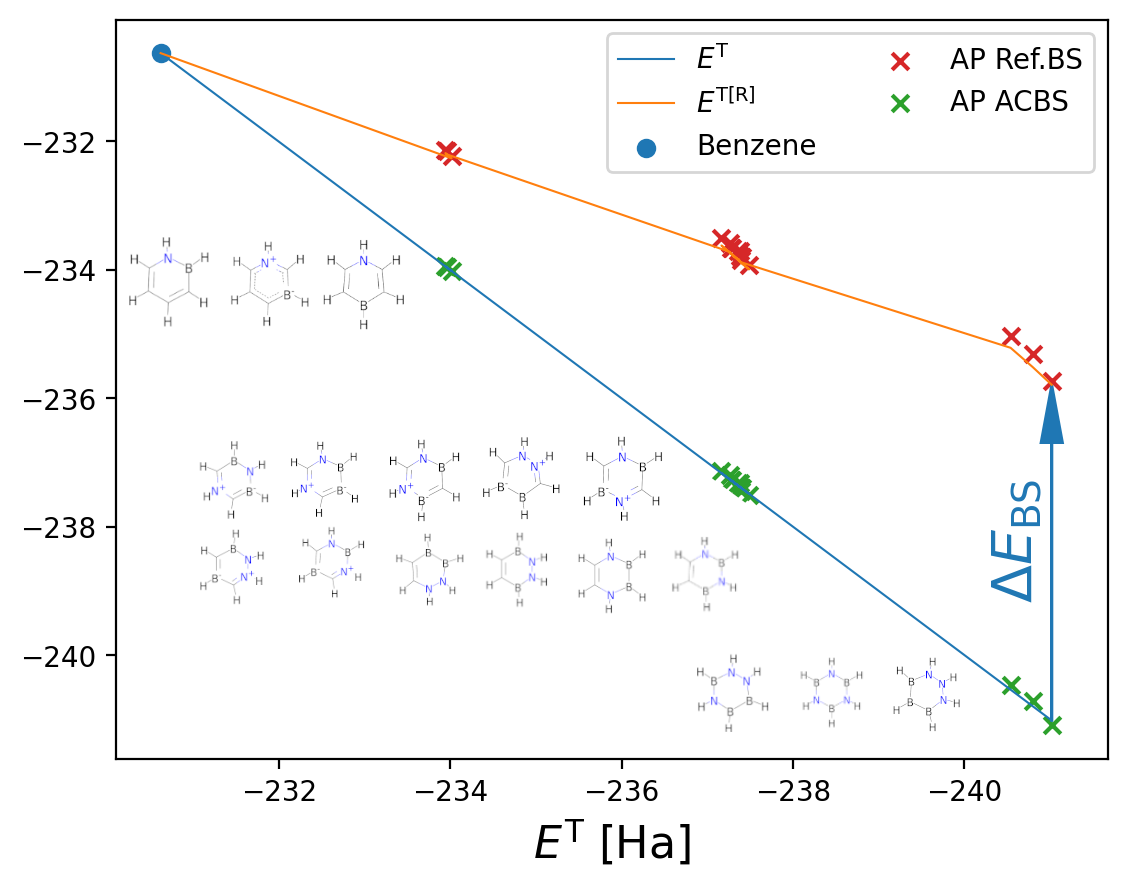}
    \caption{Alchemical B-N doping of benzene, the plot shows the true versus the second order predictions for AC-6-31G basis set, against the predictions made with the 6-31G basis set of benzene. }
    \label{fig:Benzene}
\end{figure}

\begin{table}[h!]
    \centering
    \begin{tabular}{c|c|c}
B-N substitutions & M.A.E. ACBS & M.A.E. Ref BS \\
\hline
 1 & 0.0035 &  1.8064 \\
 2 & 0.0273 &  3.6195 \\
 3 & 0.0740 &  5.4268 \\
    \end{tabular}
    \caption{MAE for the alchemical energy predictions of the BN doped benzene's mutants.}
    \label{tab:Benzene}
\end{table}

Another common application of alchemical perturbations is the BN doping of carbon aromatic systems, previous papers\cite{benzene2013Geerlings,c60_2018Geerlings,shiraogawa2023optimization,shiraogawa2022exploration,anm,Lilienfelds_bn_doped_graphene} showed how quantum alchemy can predict a combinatorially large number of doped mutants from the explicit calculation of one single reference molecule.
In figure \ref{fig:Benzene} and in table \ref{tab:Benzene} are presented the results for the second order alchemical prediction (AP2) of all 17 B-N doped mutants of benzene, calculated at the RHF level of theory, using Pople's 6-31G basis sets, the atomic coordinates are the one of the benzene's energy minimum.   
Figure \ref{fig:Benzene} shows that the alchemical predictions with the ACBS approach the true energy of the  target, while the predictions made from the reference basis set derivatives approach the energy of the mutants with the basis set of benzene ($E^{T[R]}$). The mean absolute errors for the second order predictions, listed in Table \ref{tab:Benzene} show that the error for the AC-BS can be as little as 4 mHa for a single B-N doping, while grows up to 27 mHa for double substitutions, and 74 mHa for triple substitutions, even if this is quite a large error on total energy, however constitutes less than 1\% of the total perturbation energy, and is expected that adding more terms in the series can eventually reduce this error to the desired accuracy.

\section{Conclusions}
CGTO basis sets are defined only for real elements with integer nuclear charge. In this paper I present a way to extend the basis set definition, to non-physical atoms, with non-integer nuclear charge. This is possible as long as exist a series of contiguous elements which share the same basis set contraction scheme (they possess the same number of contracted Gaussians and the same number of Gaussian primitives).

Because of the high $\Delta E_\mathrm{BS}$ some basis sets struggle to give accurate alchemical predictions of the total energy if the derivatives are calculated on the reference basis set. However if the derivatives include the alchemical consistent transformation of the atomic orbitals, as is the case of the first alchemical derivative in Equation \ref{eq:AD1}, than the alchemical perturbation series converge to the true target energy.
I showed that in section \ref{sec:num_ex} that for  alchemical perturbation with $\Delta Z =\pm 1$, such as the diatomics in Table \ref{tab:diatomics}, or a single B-N doping of benzene in Table \ref{tab:Benzene}, the error is on the milli Hartree scale. 

The Gaussian basis set interpolation reported in this paper, is analogous to the interpolation of pseudo-potential along the periodic table, showed in various works on quantum alchemy. \cite{Lilienfeld2010,Solovyeva_crystals_2016,vonLinienfeld2013firstorinciples,Chang2018,sahre2023quantum,sahre2023transferability,Griego2020MLcAP_Keith, caracas2009elasticity,marzari_94,saitta1998structural,bellaiche2002low} Stretching basis sets across the periodic table may help to smooth, not only alchemical, but also Machine Learning predictions \cite{qml_ccs_anatole2018,qml_nutshell_rupp2015,von2020exploring,qml_rupp_atomization, faber_qml_lower_DFT, christensen2019operators,weinreich2021machine,qml_properties,ceriotti2021introduction}, and in general may be useful to present a more continuous and differentiable description of the chemical compound space.

\section{Supplementary Material}
The supplementary material includes two tables: in Table \ref{Tab1_SM} are listed all the AP3 predictions obtained from the $\Delta Z = \pm 1 $ perturbation of BF,CO, and N$_2$, in Table \ref{Tab2_SM} are listed the AP2 predicted energies and errors of the B-N perturbations of benzene. 

\section{Acknowledgments}
I would like to thank professor Anatole von Lilienfeld for the nice suggestions and helpful feedback received.  

\section{Data availability statement}

The data that support the findings of this study are available from the corresponding author upon reasonable request.

\bibliography{A2_Alchemy_Cardenas,A2_Alchemy_Geerlings,A2_Alchemy_Keith,A2_Alchemy_Others,A2_Alchemy_shirogawa,A2_Alchemy_vLg,A2_BasisRef,A2_derivatives,A2_Geomopt,A2_QML,A2_software,A2_Other_works}

\newpage

\title[Supplementary materials]{Supplementary materials to the paper "Extending the definition of atomic basis sets to atoms with fractional nuclear charge"}
\author{Giorgio Domenichini}
\affiliation{University of Vienna.}

\maketitle

\begin{longtable}{ccc|ccc|cc}
\caption{\label{Tab1_SM}Third order alchemical perturbation (AP) energies of diatomic molecules using basis set of the reference, or the alchemical consistent basis set (ACBS).\label{tab:diatomics}} \\

\hline \multicolumn{1}{|c}{\textbf{Basis set}} & \multicolumn{1}{c}{\textbf{Ref. }}&  \multicolumn{1}{c|}{\textbf{Targ. }}&  \multicolumn{1}{|c}{\textbf{$E^T$}}&\multicolumn{1}{c}{\textbf{AP3 ACBS }} & \multicolumn{1}{c|}{\textbf{AP3 ref.BS}}& \multicolumn{1}{|c}{\textbf{$\Delta E$ ACBS }} & \multicolumn{1}{c}{\textbf{$\Delta E$ ref.BS}} \\
\hline 
\endfirsthead

\multicolumn{6}{c}%
{{\bfseries \tablename\ \thetable{} -- continued from previous page}} \\
\hline\multicolumn{1}{|c}{\textbf{Basis set}} & \multicolumn{1}{c}{\textbf{Ref. }}&  \multicolumn{1}{c|}{\textbf{Targ. }}&  \multicolumn{1}{|c}{\textbf{$E^T$ }}&\multicolumn{1}{c}{\textbf{AP3 ACBS }} & \multicolumn{1}{c|}{\textbf{AP3 ref.BS}}& \multicolumn{1}{|c}{\textbf{$\Delta E$ ACBS }} & \multicolumn{1}{c}{\textbf{$\Delta E$ ref.BS}} \\
\hline 
\endhead

\hline \multicolumn{6}{l}{{Continued on next page}} \\ \hline
\endfoot

\hline \hline
\endlastfoot
3-21G &  N$_2$ &    NO$^+$        &     -128.1332 &    -128.1310 &   -127.0931 &    0.0022 &     1.0401 \\
3-21G &  N$_2$ &    CN$^-$        &      -91.7316 &     -91.7354 &    -90.8583 &   -0.0038 &     0.8733 \\
  3-21G &  N$_2$ &    CO          &     -112.0879 &    -112.1286 &   -110.2232 &   -0.0406 &     1.8647 \\
  3-21G &  CO &    CF$^+$         &     -136.0534 &    -136.0556 &   -135.0153 &   -0.0021 &     1.0381 \\
3-21G &  CO &    NO$^+$           &     -128.1332 &    -128.1252 &   -127.0740 &    0.0080 &     1.0592 \\
  3-21G &  CO &    BO$^-$         &      -98.9287 &     -98.9233 &    -98.0602 &    0.0053 &     0.8685 \\
3-21G &  CO &    CN$^-$           &      -91.7316 &     -91.7309 &    -90.9029 &    0.0006 &     0.8287 \\
3-21G &  CO &    N$_2$            &     -108.3009 &    -108.2912 &   -106.4623 &    0.0097 &     1.8387 \\
3-21G &  CO &    BF               &     -123.3550 &    -123.3551 &   -121.5055 &   -0.0001 &     1.8495 \\
3-21G &  BF &   BNe$^+$           &     -151.6377 &    -151.6466 &   -150.5707 &   -0.0089 &     1.0670 \\
3-21G &  BF &   CF$^+$            &     -136.0534 &    -136.0536 &   -134.9252 &   -0.0001 &     1.1283 \\
3-21G &  BF &   BeF$^-$           &     -113.3302 &    -113.3359 &   -112.4327 &   -0.0057 &     0.8975 \\
3-21G &  BF &    BO$^-$           &      -98.9287 &     -98.9363 &    -98.1199 &   -0.0077 &     0.8088 \\
3-21G &  BF &       CO            &     -112.0879 &    -112.0907 &   -110.2067 &   -0.0027 &     1.8813 \\
3-21G &  BF &      BeNe           &     -142.0281 &    -142.0463 &   -140.0919 &   -0.0182 &     1.9363 \\
6-31G &  N$_2$ &    NO$^+$        &    -128.7977 &     -128.7971 &   -127.7504 &    0.0006 &     1.0473 \\
6-31G &  N$_2$ &    CN$^-$        &     -92.2174 &      -92.2133 &    -91.3502 &    0.0041 &     0.8672 \\
6-31G &  N$_2$ &    CO            &    -112.6616 &     -112.6696 &   -110.7722 &   -0.0080 &     1.8893 \\
6-31G &  CO &    CF$^+$           &    -136.7430 &     -136.7460 &   -135.7128 &   -0.0030 &     1.0302 \\
6-31G &  CO &    NO$^+$           &    -128.7977 &     -128.7923 &   -127.7359 &    0.0054 &     1.0618 \\
6-31G &  CO &    BO$^-$           &     -99.4492 &      -99.4464 &    -98.5853 &    0.0028 &     0.8639 \\
6-31G &  CO &    CN$^-$           &     -92.2174 &      -92.2201 &    -91.3409 &   -0.0027 &     0.8765 \\
  6-31G &  CO &    N$_2$          &    -108.8679 &     -108.8713 &   -106.9614 &   -0.0034 &     1.9065 \\
6-31G &  CO &    BF               &    -123.9888 &     -123.9932 &   -122.1201 &   -0.0044 &     1.8687 \\
6-31G &  BF &   BNe$^+$           &    -152.4083 &     -152.4127 &   -151.4068 &   -0.0044 &     1.0016 \\
6-31G &  BF &   CF$^+$            &    -136.7430 &     -136.7411 &   -135.6697 &    0.0019 &     1.0733 \\
6-31G &  BF &   BeF$^-$           &    -113.9089 &     -113.9039 &   -113.0392 &    0.0049 &     0.8697 \\
6-31G &  BF &    BO$^-$           &     -99.4492 &      -99.4523 &    -98.5550 &   -0.0031 &     0.8942 \\
6-31G &  BF &       CO            &    -112.6616 &     -112.6566 &   -110.7279 &    0.0049 &     1.9336 \\
6-31G &  BF &      BeNe           &    -142.7390 &     -142.7342 &   -140.8718 &    0.0048 &     1.8671 \\
cc-pvDZ &  N$_2$ &    NO$^+$      &    -128.9207 &     -128.9180 &   -127.8830 &    0.0027 &     1.0377 \\
cc-pvDZ &  N$_2$ &    CN$^-$      &     -92.2820 &      -92.2802 &    -91.4117 &    0.0018 &     0.8703 \\
cc-pvDZ &  N$_2$ &    CO          &    -112.7483 &     -112.7589 &   -110.8809 &   -0.0106 &     1.8674 \\
cc-pvDZ &  CO &    CF$^+$         &    -136.8473 &     -136.8490 &   -135.8296 &   -0.0017 &     1.0177 \\
cc-pvDZ &  CO &    NO$^+$         &    -128.9207 &     -128.9163 &   -127.8615 &    0.0044 &     1.0592 \\
cc-pvDZ &  CO &    BO$^-$         &     -99.5217 &      -99.5175 &    -98.6531 &    0.0041 &     0.8685 \\
cc-pvDZ &  CO &    CN$^-$         &     -92.2820 &      -92.2823 &    -91.4156 &   -0.0004 &     0.8663 \\
cc-pvDZ &  CO &    N$_2$          &    -108.9554 &     -108.9605 &   -107.0683 &   -0.0051 &     1.8871 \\
cc-pvDZ &  CO &    BF             &    -124.0575 &     -124.0605 &   -122.2099 &   -0.0031 &     1.8475 \\
cc-pvDZ &  BF &   BNe$^+$         &    -152.4629 &     -152.4681 &   -151.4674 &   -0.0052 &     0.9955 \\
cc-pvDZ &  BF &   CF$^+$          &    -136.8473 &     -136.8434 &   -135.7591 &    0.0038 &     1.0882 \\
cc-pvDZ &  BF &   BeF$^-$         &    -113.9656 &     -113.9620 &   -113.1076 &    0.0036 &     0.8580 \\
cc-pvDZ &  BF &    BO$^-$         &     -99.5217 &      -99.5250 &    -98.6259 &   -0.0034 &     0.8958 \\
cc-pvDZ &  BF &       CO          &    -112.7483 &     -112.7363 &   -110.8019 &    0.0120 &     1.9464 \\
cc-pvDZ &  BF &      BeNe         &    -142.7729 &     -142.7520 &   -140.9158 &    0.0210 &     1.8571 \\
\end{longtable} 

\begin{table}
\caption{ \label{Tab2_SM} Second order alchemical perturbation energies of B-N doped benzene's mutants, using basis set of the reference, or the alchemical consistent basis set.}
\begin{tabular}{l|cccc}
\toprule
Substitution &    AP2 ref.BS&  $E^{T[R]}$ &  AP2 ACBS &   $E^T$  \\
\hline
 1N 2B & -232.2207  &  -232.2638   &   -234.0194  &  -234.0179   \\
1N 3B & -232.1337  &  -232.1833   &   -233.9415  &  -233.9453   \\
1N 4B & -232.1521  &  -232.2011   &   -233.9575  &  -233.9625   \\
1,2 N 3,4 B& -233.5743  &  -233.7186   &   -237.1966  &  -237.2612   \\
1,2 N 3,5 B & -233.6614  &  -233.7664   &   -237.2745  &  -237.2895   \\
1,2 N 3,6 B & -233.7116  &  -233.8600   &   -237.3205  &  -237.3774   \\
1,2 N 4,5 B & -233.5057  &  -233.6205   &   -237.1347  &  -237.1658   \\
1,3 N 2,4 B & -233.9224  &  -233.9920   &   -237.5080  &  -237.4881   \\
1,3 N 2,5 B & -233.7986  &  -233.8982   &   -237.3983  &  -237.4021   \\
1,3 N 4,5 B & -233.6614  &  -233.7425   &   -237.2745  &  -237.2776   \\
1,3 N 4,6 B  & -233.8538  &  -233.8843   &   -237.4461  &  -237.3925   \\
1,4 N 2,3 B  & -233.7116  &  -233.8326   &   -237.3205  &  -237.3640   \\
1,4 N 2,5 B & -233.7802  &  -233.8728   &   -237.3824  &  -237.3844   \\
1,4 N 2,6 B  & -233.7986  &  -233.8799   &   -237.3983  &  -237.3915   \\
1,2,3 N 4,5,6 B & -235.0334  &  -235.2137   &   -240.4677  &  -240.5377   \\
1,2,4 N 3,5,6 B & -235.3078  &  -235.5129   &   -240.7153  &  -240.7952   \\
1,3,5 N 2,4,6 B & -235.7295  &  -235.7904   &   -241.0905  &  -241.0183   \\
\end{tabular}
\end{table}

\end{document}